\documentclass[journal,comsoc]{IEEEtran}

\usepackage{booktabs} 
\usepackage{caption}
\usepackage{subcaption}
\usepackage{multirow}
\usepackage{longtable}
\usepackage{listings}
\usepackage{longtable}
\usepackage{xcolor,colortbl}
\definecolor{maroon}{cmyk}{0,0.87,0.68,0.32}
\definecolor{orange}{rgb}{1,0.5,0}
\definecolor{green}{rgb}{70,130,180}
\usepackage{dirtytalk}
\usepackage{tikz}

\usepackage{array}
\usepackage[labelfont=bf]{caption}
\usepackage{xcolor} \usepackage{multirow, ltablex, colortbl, makecell}

\usepackage{ragged2e}
\usepackage{lipsum}
\usepackage{amssymb}
\usepackage{pifont}
\newcommand{\cmark}{\ding{51}}%
\newcommand{\xmark}{\ding{55}}%

\ifCLASSOPTIONcompsoc

  \usepackage[nocompress]{cite}
\else
  \usepackage{cite}
\fi

\ifCLASSINFOpdf

\fi

\usepackage{amsmath}

\usepackage{comment}

\usepackage{algorithmicx}
\usepackage[ruled,vlined]{algorithm2e}

\begin{document}

\title{Using Harmonics for Low-Cost Jamming}
\author{Vasilis Ieropoulos}
\author{
Vasilis Ieropoulos \textsuperscript{1,*},Eirini Anthi\textsuperscript{2}\thanks{*Corresponding author: ieropoulosv@cardiff.ac.uk,}

 \textsuperscript{1}Cardiff University, School of Computer Science \& Informatics, Cardiff, UK 
}

\maketitle

\IEEEtitleabstractindextext{%
\begin{abstract}

The digitalisation of the modern schooling system has led to multiple schools and organisations buying similar hardware. Electronic equipment like wireless microphones, projectors, touchscreen displays etc., have been almost standardised with a few well-known brands leading the market. This has led to the adoption of common frequency ranges between brands with many sticking between 600-670 MHz. The popularity of low-cost computing devices like the Raspberry Pi which has been used in a plethora of applications has also taken the path of being used as low-cost transmitters. There have been many implementations where the Raspberry Pi has been used as the target device but few cases where the PI is the actual threat. In this paper, we explore the use of the Raspberry Pi as a stealth radio frequency jamming device to disable wireless conference microphones. Harmonics were used to achieve frequencies outside the Pi's transmission frequency by taking advantage of its unfiltered transmission.

\end{abstract}

\begin{IEEEkeywords}
Internet of Things (IoT), Open-Source, Radio Frequency, Software Defined Radio 
\end{IEEEkeywords}}

\IEEEdisplaynontitleabstractindextext

\IEEEpeerreviewmaketitle

\ifCLASSOPTIONcompsoc
\IEEEraisesectionheading{\section{Introduction}\label{sec:introduction}}
\else
\section{Introduction}
\label{sec:introduction}
\fi

\IEEEPARstart{T} he uses of wireless technologies has risen substantially in the past few years, and the adoption of wireless devices with integrated sensors has become ever more prevalent in the modern age\cite{kim2006introduction}. Many industries have become dependent on wireless devices for monitoring buildings, infrastructure, health, organizations etc\cite{chevrollier2005use}. A large portion of these devices are part of an interconnected network, with many of them being portable or located in places where hard-wiring is not an option. Therefore, it is unavoidable that these devices are dependent on radio communication for the exchange of data. The knowledge on how to perform radiofrequency attacks has usually been confined to a small group of people with very specific knowledge on the subject, as well as the use of specialized hardware\cite{chu2018penetration}. This makes it difficult to assess security vulnerabilities that are present in RF technology, as the time and financial constraints of third-party contractors hired to assess these are limited\cite{chu2018penetration}. 
This paper focuses on the uses of a low-cost single-board computer like the Raspberry Pi to perform a stealthy jamming attack on a wireless conference microphone. The paper emphasizes the ease of use of open-source software to achieve this, as well as the low cost of the overall platform. Furthermore, the use of Harmonics is used to overcome the physical limitations of the device itself by using basic physics to exploit the otherwise useless harmonic emissions of unfiltered signal transmissions. Using a combination of low-cost hardware and the exploitation of the unfiltered signals produced by the device allows for simple, yet effective RF penetration.  

This paper is structured as follows: Section \ref{sec:Related Work} discusses related work. Section \ref{sec:Testing and Methodology} describes the methodology and testing of the experiment. Section \ref{sec:Mitigation} covers mitigation strategies, while Section \ref{sec:Future work} outlines plans for future work and improvements.

\section{Related Work}\label{sec:Related Work}
The field of Radio Frequency penetration testing is complex but interesting and is usually done using very expensive equipment. In the case of Jean-Michel Picod et al.,\cite{picod2014bringing} an Ettus USRP B210 board was used to perform an RF vulnerability assessment of different wireless devices like Zigbee sensors. The use of such a device, although it is effective, requires plenty of training to use, as well as the high price tag compared to more low-cost solutions makes it less appealing to the average hobbyist. The Raspberry Pi has been used in penetration testing mostly for network-related attacks as demonstrated by Maryna Yevdokymenko et al.\cite{8246375} which used the Pi along with Kali Linux to perform network penetration tests. The size of the Pi makes it perfect for stealth attacks as its low profile makes it hard to notice, and its low power consumption allows it to be run using only a power bank. The Pi is no stranger to being used as a wireless network tool as well as demonstrated by Dr Chafic BouSaba et al.\cite{asee_peer_27206} who also attached a 16×2 Character LCD and Keypad Kit. This allowed information to be displayed to the user directly from the Pi without the need for a remote connection. Hence, this opens up the Pi for automated scripts that can be executed without physically accessing it. Therefore, there exists the possibility of using the Raspberry Pi as a portable, standalone and stealthy RF penetration tool. 

\begin{table}[h]
\centering

\scalebox{0.9}{

\begin{tabular}{|l|l|l|ll}
\cline{1-3}
\multicolumn{1}{|c|}{\textbf{Paper}} & \multicolumn{1}{c|}{\textbf{Equipment}} & \multicolumn{1}{c|}{\textbf{Use of external equipment}} &  &  \\ \cline{1-3}
\cite{picod2014bringing} & USRP                 & Yes &  &  \\ \cline{1-3}
\cite{8246375}           & Rpi, Usb dongle      & Yes &  &  \\ \cline{1-3}
\cite{asee_peer_27206} & Pi, External Screen & Yes &  &  \\ \cline{1-3}
This Paper                                & Pi                   & No  &  &  \\ \cline{1-3}
\end{tabular}
}
\end{table}

\subsection{Harmonics}
Harmonics are produced by every oscillating wave, with each level of the harmonic being a different frequency. Harmonics have a reduction of bandwidth with every iteration, while also reducing their power\cite{harmonics}. They can be calculated using the following formula:

\begin{equation}
    \mathbf{Harmonic = 2 \cdot  \pi \cdot   \textit{f} \cdot   n}
\end{equation}

where \textit{f} is the frequency of the carrier and n is the number of the harmonic.

The advantage of using harmonics, in this case, is that they can be used to send signals outside the 500MHz range, assuming no filter is attached. Furthermore, they can cause the capacitance level of power factor capacitors to fluctuate and the inductance of the power supply to become unstable, causing the device or neighbouring devices to behave irregularly\cite{wi-fi-radio_modulation_2021}. Spurious emissions are also one of the main reasons for using harmonics, as they can be used to create a wide range of frequencies close to the fundamental frequency of oscillation, which can cause interference.
Many of these spurious emissions are generated by Low Noise Amplifier embedded in the devices, which are used to amplify the low powered signal transmitted from the device. 


\subsection{rpitx}
\begin{figure}[h!]
    \centering
    \includegraphics[scale=0.4]{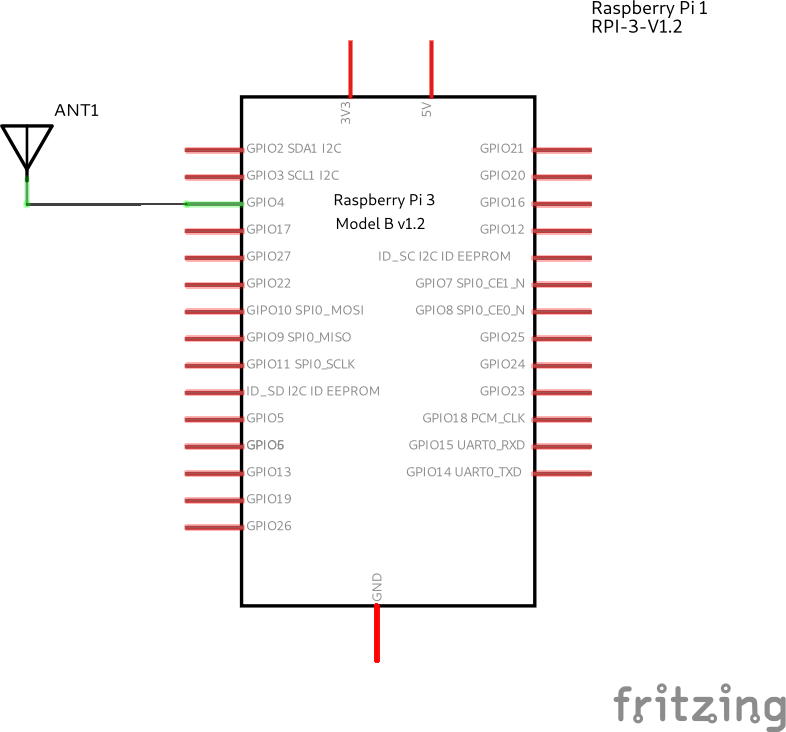}
    \caption{Raspberry PI schematic}
    \label{fig:Raspberry}
\end{figure}

rpitx is a general radio frequency transmitter created by Evariste Courjaud(F5OEO)\cite{f5oeo_2019} for the Raspberry Pi which doesn't require any other hardware to operate, but a filter is recommended to avoid unwanted interference. It can handle frequencies from 5 kHz up to 1500 MHz. It uses the GPIO pins of the Raspberry Pi to control the frequency and the power of the transmitter. Specifically, it uses GPIO4 or pin 7 of the GPIO header. The signal being transmitted is unfiltered which means that a bandpass filter is required for it to be used without causing interference. It allows the Pi to transmit using multiple modulations types such as AM, FM, and SSB. Observing figure \ref{fig:Raspberry} it demonstrates where the short piece of wire can be installed on the GPIO header. 

The early versions of the software only allowed for transmissions up to 500MHz, which made testing out the hypothesis difficult. This meant a different approach had to be taken for a more out-of-the-box solution. Hence, using harmonics made sense as it would allow the transmission of signals outside the 500Mhz range along with the Raspberry Pi's unfiltered signal transition. Even though sending unfiltered RF is dangerous, the low power of the Raspberry Pi allows us to harness the harmonic transmissions in a meaningful way in a controlled environment.

\section{Testing and Methodology}\label{sec:Testing and Methodology}

\subsection{Methodology} 

\begin{figure}[ht!]
    \centering
    \includegraphics[scale = 0.1]{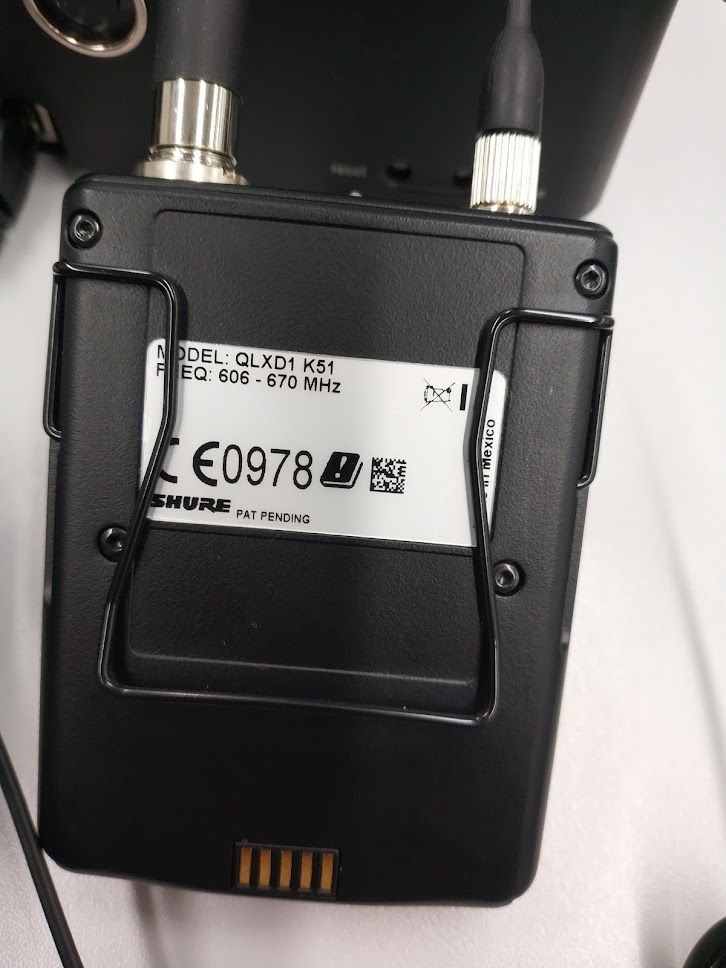}
    \caption{Microphone Backplate}
    \label{fig:mic}
\end{figure}

The model number and the frequency needed to be identified. Photographic material was taken of the device, identifying it as a "SHURE QLXD1" with a frequency range between 606-670 MHz. This was clearly labelled on the back of the device, as shown in figure\ref{fig:mic}.

\begin{figure}[h!]
    \centering
    \includegraphics[scale = 0.1]{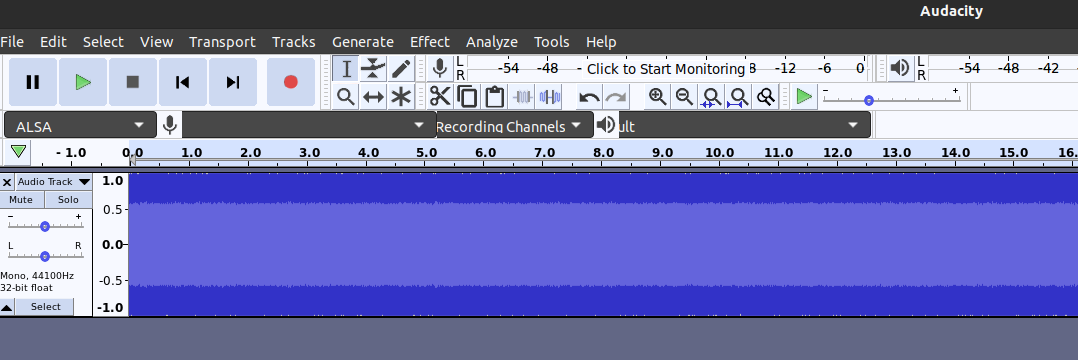}
    \caption{White Noise Generation}
    \label{fig:noise}
\end{figure}

The next step was to identify which was the frequency of operation of the device by scanning the frequency range using GQRX and the SDR. The frequency was identified as 614 MHz. We continued monitoring the frequency before and after the device was turned off and on during use. It was identified that there seemed to be an initial handshake packet that was used to establish a connection between the microphone and the wireless receiver. This was also used for establishing frequency hopping between the two devices. As demonstrated in figure \ref{fig:noise} white noise was created using audacity as a sample file. This allowed us to create the widest possible signal possible, which would allow the transmission to cover a large portion of the band equally.

This potentially could give us the best results. While monitoring the signal, it was observed that the transmission was using FM modulation, hence the sample file had to be converted to accommodate for this. This was done using the following command:
\begin{lstlisting}

./pifm sample.wav fm.ft

\end{lstlisting}

This converted the WAV file into an FM-modulated sound file, which allows rpitx to transmit it in this fashion. 
\subsection{Testing}
Since the microphone was operating at 614Mhz, which was outside the operating frequencies of the software. An out-of-the-box solution had to be used. Hence, the Raspberry Pi was set up to operate at 304 MHz which would create a second harmonic at 614 MHz, allowing us to attack the frequency of operation. 
To test our hypothesis, we started transmitting the sound file while the microphone was in use. This did not affect the microphone, as the handshake signal which was initiated in the beginning also initiated some form of frequency hopping, making it difficult for the Raspberry Pi to stop or interfere with the transmission. 
We ran the test a second time but in this case, started transmitting before the microphone was connected to the receiver. This caused the microphone to not be able to connect to the receiver as long as we had the Raspberry Pi transmitting. The following was also attempted but resulted in the microphone never connecting:
\begin{enumerate}
    \item Turning both devices on and off again 
    \item Hard resetting the devices by removing the batteries
\end{enumerate}

To further expand the testing, the transmission was also done on different frequencies which produced harmonics at different increments but still on the frequency of operation of the microphone. As demonstrated in Table \ref{tab:harmonics} the experiment was successful until the fifth harmonic, after that, the reduction in power and bandwidth did not allow for any further results.
\begin{table}[h!]
\centering
\begin{tabular}{|l|l|l|}
\hline
\textbf{Frequency} & \textbf{Harmonic} & \textbf{Result} \\ \hline
304 MHz            & 2nd               & \cmark          \\ \hline
204.7 MHz          & 3rd               & \cmark          \\ \hline
153.5 MHz          & 4th               & \cmark          \\ \hline
102.4 MHz          & 5th               & \xmark          \\ \hline
\end{tabular}
\caption{Harmonics Tested}
\label{tab:harmonics}
\end{table}

\section{Mitigation Strategies} \label{sec:Mitigation}

Although this is quite a complex topic with a simple implementation, the following are some of the mitigation strategies that can be used to mitigate the effects of jamming attacks. One of the main mitigation strategies is to use a filter to avoid interference\cite{misra2018development}. This can be done by using a bandpass filter to filter out the unwanted frequencies, as well as using an attenuator to attenuate the signal\cite{misra2018development}. Even though in this case the Raspberry Pi managed to stop the microphone from initiating the handshake signal, thus stopping it from working, there are solutions to try to reduce the chance of this happening. For example, the charging dock could use the physical pins on the back of the microphone transmitter to initialise the handshake signal directly, instead of relying on the wireless capabilities of the transmitter\cite{peeters2001synchronous}. This would eliminate the chance of an outside source interfering with the transmission.

\section{Conclusion} \label{sec:conclusion}

The substantial increase in wireless devices over the past decade has made them an attractive target for malicious threat actors. The use of wireless devices has also made it easier for malicious actors to target them, as they can be easily located by using a wireless scanner such as an SDR. This makes it easier for hackers to target the devices and then exploit them for their own purposes. The intimidating part about pure radio frequency attacks is the difficulty of detecting that an attack is happening, especially in the form of sniffing. The Raspberry Pi has proved itself to be a very reliable device for wireless attacks, as well as being used for more sophisticated attacks such as replay attacks and packet manipulation. Its low cost makes it an excellent solution for hobbyists and tinkerers alike. In our scenario, the Raspberry Pi was used to jam the communications of a wireless conference microphone, which could disrupt the talk or conference in motion. This could be expanded into eavesdropping on confidential or private meetings which use such devices, which as demonstrated are transmitting unencrypted voice transmissions. The Pi can be concealed easily due to its small size, making it a potential threat in the next RF attack.

\section{Future Work} \label{sec:Future work}

The experiment does not come without its limitations. The newer versions of the RPTIX software allow for transmissions of up to 1.5 GHz, which gives the user more freedom to explore the vulnerabilities in the spectrum and to use the device for more than just jamming. More sophisticated attacks using a cheap software-defined radio could be used for replaying packets back to devices, as well as opening up the possibility of sniffing and packet manipulation. At the time of writing, the Raspberry Pi 4 is available and might be more effective in these types of attacks. Furthermore, some boards take advantage of the GPIO pin, giving the user the ability to connect an antenna using an SMA connector with embedded filtering on the board. One such example is the CaribouLite RPi HAT\cite{kocher_2021} which can tune up to 6GHz, this would allow the Pi to be used in more high-frequency scenarios like Wi-Fi\cite{Ost_2018}, Bluetooth\cite{scientific}, Zigbee\cite{FARAHANI200825} etc.

\bibliographystyle{IEEEtran}
\bibliography{ref}

\end{document}